\documentclass{llncs}
\usepackage{makeidx}  
\usepackage{epsfig}

\begin{document}
\pagestyle{empty}  
\bibliographystyle{splncs}
\title{Critical Missing Equation of Quantum Physics\\for Understanding Atomic Structures}
\author{Xiaofei Huang}
\institute{School of Information Science and Technology\\Tsinghua University, Beijing, China, 100084 \\
\email{HuangZeng@yahoo.com}}

\maketitle

\begin{abstract}
This paper presents an optimization approach 
	to explain why and how a quantum system evolves from an arbitrary initial state to a stationary state, 
	satisfying the time-independent Schr\"{o}dinger equation. 
It also points out the inaccuracy of this equation, which is critial important in quantum mechanics and quantum chemistry,
	due to a fundamental flaw in it conflicting with the physical reality. 
The some directions are suggested on how to modify the equation to fix the problem\footnote{Presented in the 12th international conference on Mathematical Results in Quantum Mechanics~(QMATH12), 
	Humboldt University of Berlin, September, 2013.}.
\end{abstract}

\section{Introduction}

Our current understanding of electron motions in an atom 
	is just like the Kepler era at understanding planetary motions in our solar system. 
We know the mathematical formula for describing the electron orbit (the electron cloud to be more accurate). 
However, we do not know why it must satisfy the formula and how an electron moves from one orbit to another. 
This paper postulates that the answers to these two critical questions may lie in a global optimization algorithm. 
The electron orbits are the equilibrium points of the algorithm and the orbit jumping can be understood 
	as the energy minimization process defined by the algorithm. 
It is desirable for nature to deploy a global optimization process at constructing atoms and molecules for their consistency and stability. 

Specifically, nature needs to ensure that atoms and molecules can evolve from arbitrary initial states to their lowest energy states, 
	even when their energy landscapes are full of local minima. 
Such a point of view can serve as a theoretical basis for us to understand why atoms and molecules in nature are stable 
	and have definite and unchangeable structures and properties. 
Otherwise, if those basic constituents of nature can easily get stuck at arbitrary states other than their ground states, 
	then there is no way to ensure their stability and consistency. 
Consequently, the universe will fall into chaos and it is impossible to have everything including life 
	because all living organisms are made of the large size molecules, called proteins.

\section{Optimization Approach to Quantum Mechanics}

The global optimization algorithm for constructing atoms and molecules can be described as the following equation:
\begin{equation}
-\hbar\frac{\partial \Psi}{\partial t}  = H\Psi -\langle H \rangle \Psi,                                          
\label{Eq1}
\end{equation}
where $\hbar$ is the reduced Planck constant, $\Psi(x_1, x_2, \ldots, x_n,t)$, or simply denoted as $\Psi(x,t)$, 
	is the wavefunction describing the state of a quantum system of $n$ particles, 
$H(x_1, x_2, \ldots, x_n,t)$ is the Hamiltonian operator corresponding to the total energy of the system at time $t$, 
	$x_i$ is the spatial position of the particle $i$ in the system, and $\langle H \rangle$ is the expected value of $H$, 
	defined as $\langle H \rangle= (\Psi, H\Psi)$. 

The above equation is so called the quantum optimization equation. 
It can be derived from a global optimization algorithm (see \cite{Huang2053}). 
In mathematics, we can prove that the equation~(\ref{Eq1}) guarantees to minimize the system's total energy $\langle H \rangle$ (see Appendix A).
It also defines a unitary transformation for $\Psi(x,t)$  such that $||\Psi||^2$ remains unchanged with time (see Appendix B for proof).
Otherwise, it would not make any sense in physics.

The quantum optimization equation takes the dual form of a key equation in quantum mechanics, called the time-dependent Schr\"{o}dinger equation:
\begin{equation}
i \hbar \frac{\partial \Psi}{\partial t}  = H \Psi \ . 
\label{SE}
\end{equation}
The quantum optimization equation can be obtained simply by replacing the imaginary number $i$ by the number $-1$ in the time-dependent Schr\"{o}dinger equation. 
The second term on the right side of the quantum optimization equation, $\langle H \rangle \Psi$, 
	is solely for ensuring that the equation defines a unitary transformation for the wavefunction $\Psi$. 
Note that the time-dependent Schr\"{o}dinger equation also defines a unitary transformation for $\Psi$.

The two equations are complementary to each other. 
The quantum optimization equation~(\ref{Eq1}) describes the dynamics for an energy-dissipative many-body system 
	while the time-dependent Schr\"{o}dinger equation~(\ref{SE}) describes that for an energy-conservative one. 

Most importantly, we can see that the quantum optimization equation at any equilibrium point 
	falls back to the stationary Schr\"{o}dinger equation, 
\begin{equation}
H \Psi=\langle H \rangle \Psi \ . 
\label{stationary_SE}
\end{equation}
That is, assume that the dynamics of a many-body system is governed by the quantum optimization equation~(\ref{Eq1}). 
Starting from any initial state, its total energy $\langle H \rangle$ always decreases as the time elapses 
	until it reaches an equilibrium state satisfying the stationary Schr\"{o}dinger equation~(\ref{SE}), called a stationary state in physics. 
Most often the system converges to the lowest energy equilibrium state, called the ground state in physics, because other equilibrium states are not stable. 
Any small disturbing force, such as quantum vacuum fluctuation or background radiation, 
	will shake the system out of a unstable, stationary state of a higher energy level and it will converge to the stable ground state eventually.

Assume that the Hamiltonian is not time-dependent, i.e., it has the form of $H(x)$ instead of $H(x,t)$. 
Based on the spectrum theorem in mathematics, the time-evolution of the state of a system governed by the quantum optimization equation~(\ref{Eq1}) is
\begin{equation}
\Psi(x, t)=e^{-H(x)t/ \hbar}\Psi(x, 0)=\frac{1}{z(t)}\left(\sum^{m}_{i=1} c_i e^{-E_i t/ \hbar}\Phi_i\right) \ , 
\label{dual_SE1}
\end{equation}
where $z(t)$ is a normalization factor such that $||\Psi(x, t)||^2=1$. 
$E_1,E_2,\ldots,E_m$ are the eigenvalues of $H(x)$ satisfying
\[ E_1  < E_2< \cdots < E_ m \ . \]
$\Phi_1,\Phi_2, \ldots, \Phi_m$ are the associated eigenvectors. 
The initial state $\Psi(x,0)$ of the system can be represented as
\[ \Psi(x, 0)=c_1 \Phi_1+c_2 \Phi_2+ \cdots +c_m \Phi_m \ . \]

Contrary to that, the time-evolution governed by the Schr\"{o}dinger equation~(\ref{SE}) is
\[ \Psi(x, t)=e^{-iH(x)t/ \hbar}\Psi(x, 0)=\sum^{m}_{i=1} c_i e^{-iE_i t/ \hbar}\Phi_i \ , \]

Assume that the initial state $\Psi(x, 0)$ has a component of the ground state $\Phi_1$, i.e., $c_1 \not= 0$.
According to Eq.~(\ref{dual_SE1}), we have
\[ \lim_{t \rightarrow \infty} \Psi(x, t)=\Phi_1 \ . \]
The global minimum $\Phi_1$, thus, found by the quantum optimization equation~(\ref{Eq1}). 
In other words, the equation~(\ref{Eq1}) has the power to find the global optimum to minimize the system's total energy $\langle H(x) \rangle$.

In particular, for a $n$-particle quantum system, 
	if we approximate $\Psi$ by factorizing it into the Cartesian product of the single particle states $\Psi_i(x_i,t)$ as follows 
\[ \Psi(x_1,x_2, \ldots, x_n, t)=\Psi_1(x_1,t)\Psi_1(x_1,t) \cdots \Psi_n(x_n,t) \ , \] 
the quantum optimization equation~(\ref{Eq1}) is reduced to
\begin{equation}
-\hbar \frac{\partial \Psi_i(x_i,t)}{\partial t}  = (H_i(x_i,t) -\langle H_i (x_i,t) \rangle) \Psi_i(x_i,t)\ , \quad \mbox{for $i=1,2, \ldots, n$}, 
\label{Eq2}
\end{equation}
where $H_i(x_i,t)$  is the local energy for the particle $i$ defined as 
\[ H_i(x_i)=(\Psi_{-i}, H(x_1, x_2,\ldots,x_n)\Psi_{-i}) \ , \]
and $\Psi_{-i}$ is the Cartesian product of the states of all the particles excluding the particle $i$ itself, 
	i.e., 
\[\Psi_{-i} =\prod_{j,j\not=i} \Psi_j \ . \] 

Just like the original one, the equation~(\ref{Eq2}) defines a unitary transformation for $\Psi_i(x_i,t)$
	and is guaranteed to minimize the system's total energy $\langle H \rangle$ (the proofs are provided in Appendix C and Appendix D). 
However, due to the approximation of the wave function by the factorization,
	the performance of this optimization equation~(\ref{Eq2}) is degraded from the original one~(\ref{Eq1})
	because it can have more than one stable equilibrium and finds the lowest energy one among them is not guaranteed.
	
The equation~(\ref{Eq2}) can also be viewed as a gradient descent algorithm for minimizing $\langle H \rangle$ 
	with respect to $\Psi_1,\Psi_2,\ldots,\Psi_n$, subject to $||\Psi_i||^2=1$, for $1 \le i \le n$. 
For a classical multivariate function $E(x)$, 
	it shares exactly the same global optimum with $\langle E \rangle$ .
Hence, minimizing $E$ with respect to $x_1, x_2, \ldots, x_n$ is same as minimizing $\langle E \rangle$ 
	with respect to $\Psi_1(x_1),\Psi_2(x_2), \ldots, \Psi_n(x_n)$. 
However, an algorithm based on the former may suffer terribly with the local minimum problem.
It may get stuck into one local minimum or another, 
	sensitive to initial condition and perturbations.
Contrary to that, an algorithm based on the later, such as the quantum optimization equation~(\ref{Eq2}), 
	can greatly alleviate the local minimum problem.
The mathematical analysis shows that the global optimization power of the equation~(\ref{Eq2}) comes from 
	smoothing out the original energy landscape $E(x)$ by making the soft-decision $\Psi_i(x_i)$ at assigning decision variable $x_i$.
The energy landscape smoothing can remove local minima, and it can eliminate all of them for some cases.

For the convenience of computer implementations, the equation~(\ref{Eq2}) can be modified as
\begin{equation}
\Psi_i(x_i,t+\Delta t)  = \frac{1} {z_i(t)} e^{-H_i(x_i)\Delta t/ \hbar}\Psi_i(x_i,t) \quad \mbox{for $1\le i \le n$.}
\label{Eq3}
\end{equation}

The optimization equation~(\ref{Eq3}) is also guaranteed to minimize the system's total energy $\langle H \rangle$ (see Appendix E for the proof). 
Here the constant $\hbar$ just controls the convergence rate which can be replaced by any positive value. 
Both Eq.~(\ref{Eq2}) and Eq.~(\ref{Eq3}) share the same equilibria defined by a modified version of the time-independent Schr\"{o}dinger equation as follows
\begin{equation}
H_i \Psi_i=\langle H_i \rangle \Psi_i \quad \mbox{for $1\le i \le n$ } \ . 
\label{Eq4}
\end{equation}

The quantum optimization equations in the reduced forms (\ref{Eq2}) and (\ref{Eq3}) can be used 
	for computing both the ground-state as well as excited states of a quantum system. 
Different from many existing methods, it is guaranteed to converge, simple in computation, 
	easy to understand and visualize, and doesn't suffer many restrictions and limitations of its predecessors. 

If we approximate $\Psi(x, t)$ as the Cartesian product of the binary particle states $\Psi_{i,j}(x_i,x_j,t)$ as follows 
\[ \Psi(x, t)=\prod_{(i,j) \in E} \Psi_{i,j} (x_i,x_j,t) \ , \]
the quantum optimization equation~(\ref{Eq1}) is reduced to
\begin{equation}
-\hbar \frac{\partial \Psi_{i,j}(x_i,x_j,t)}{\partial t}  =(H_{i,j} -\langle H_{i,j} \rangle) \Psi_{i,j}(x_i,x_j,t), \quad \mbox{for $(i,j) \in E$},
\label{Eq5}
\end{equation}
where $H_{i,j}$  is the local energy experienced by the particle $i$ and the particle $j$ defined as 
\[ H_{i,j}=(\Psi_{-i,j}, H(x_1, x_2, \ldots,x_n)\Psi_{-i,j}) \ , \]
and
\[\Psi_{-i,j} =\prod_{(i^{'}, j^{'}) \in E, (i^{'}, j^{'}) \not= (i,j)} \Psi_{(i^{'}, j^{'})} \ . \] 

The equilibrium of the equation~(\ref{Eq5}) is defined as
\begin{equation}
H_{i,j}\Psi_{i,j}=\langle H_{i,j} \rangle \Psi_{i,j} \ . 
\label{Eq6}
\end{equation}

\section{A Fundamental Flaw in the Schr\"{o}dinger Equation}

The investigation in the previous section 
	tells us that any quantum system with stationary states governed 
	by the time-independent Schr\"{o}dinger equation~(\ref{stationary_SE}), 
	should have the ground state as the only stable stationary state. 
However, in nature, some large molecules do have multiple stable sub-optimal configurations. 
That fact conflicts with our current quantum theory which reveals a fundamental flaw in it. 
Therefore, we can conclude that our current theory is not accurate at describing the physical reality. 
Furthermore, from the computational point of view, 
	defining the stationary states of a quantum system based on the time-independent Schr\"{o}dinger equation~(\ref{stationary_SE}) does not make sense 
	because it requires computing resources grow exponentially with the number of the particles in the system. 
It is a brute-force way to solve a fundamental problem in computation.

However, it is not the case for the other two equations~(\ref{Eq4}) and ~(\ref{Eq6})
	for defining the stationary states for a quantum system.
Those two equations can have multiple stable stationary states and do not suffer the exponential complexity problem.
Those two are based on two different factorizations of the wavefunction $\Psi(x,t)$.
In general, to address the exponential complexity issue in the original time-independent Schr\"{o}dinger equation~(\ref{stationary_SE}),
	we need to factorize the wave function into component functions of limited orders (unary, binary, or $n$-ary).
However, it is an open question on how to factorize the wave function to define an equation 
	for describing the stationary states of a quantum system that agrees with the physicsl reality.

\section{Conclusions}
This paper postulates that nature may deploy a global optimization algorithm at contructing atoms and molecules
	for their stability and consistency.
It has different versions depending on how the wavefunction $\Psi(x,t)$ is factorized.
The version of the algorithm without factorizing $\Psi(x,t)$ always converge to the equilibria satisfying the classical time-independent Schr\"{o}dinger equation.

This paper also points out the inaccuracy of the time-independent Schr\"{o}dinger equation at descibing the stationary states of quantum systems.
It has a single, stable stationary state conflicting with the physical reality that some large size molecules have multiple stable configurations of different energy levels in nature.
It also has the exponential complexity issue 
	which does not make sense from a computational point of view.
The different versions of the modification for the equation based on the different factorizations of the wave function $\Psi(x,t)$ 
	have also been proposed in this paper to address this problem.

\section{Appendix A}
The quantum optimization equation~(\ref{Eq1}) minimizes the system's total energy $\langle H \rangle$ (the expected value of $H$).

\begin{proof}
\begin{eqnarray*}
\frac{d}{dt} \langle H \rangle & = & \frac{d}{dt} (\Psi, H\Psi) \\
& = & \left(\frac{\partial \Psi}{\partial t}, H \Psi \right) + \left(\Psi, H \frac{\partial \Psi}{\partial t} \right)  \\
& = & \left( -\frac{1}{\hbar} (H -\langle H \rangle) \Psi, H\Psi \right) + \left(\Psi, -\frac{1}{\hbar}H(H -\langle H \rangle) \Psi \right) \\
& = &  -\frac{2}{\hbar} \langle H^2 \rangle - \frac{2}{\hbar} \langle H \rangle^2 \\
& \le & 0
\end{eqnarray*}	
The equality at the last step of the above proof holds true if and only if 
	$\Psi$ satisfies the following equation:
\begin{equation} 
H \Psi = \langle H \rangle \Psi. 
\label{a_1}
\end{equation}
In this case, 
	the equation reaches an equilibrium because $\partial \Psi(x,t) / \partial t = 0$. 

The equation~(\ref{a_1}) is exactly the time-independent Schr\"{o}dinger equation~(\ref{stationary_SE}) in quantum mechanics
	when we note that $\lambda = \langle H \rangle$.
\end{proof}

\section{Appendix B}
The quantum optimization equation~(\ref{Eq1}) defines a unitary transformation for $\Psi(x, t)$.
\begin{proof}
\begin{eqnarray*}
\frac{d}{dt} |\Psi|^2 & = & \frac{d}{dt} ( \Psi, \Psi )\\
& = & \left(\frac{\partial \Psi}{\partial t}, \Psi \right) + \left(\Psi, \frac{\partial \Psi}{\partial t} \right)  \\
& = & \left( -\frac{1}{\hbar} (H -\langle H \rangle) \Psi, \Psi \right) + \left(\Psi, -\frac{1}{\hbar}(H -\langle H \rangle) \Psi \right) \\
& = &  -\frac{2}{\hbar} (\Psi, H \Psi) - \frac{2}{\hbar} (\Psi, \langle H \rangle \Psi) \\
& = &  -\frac{2}{\hbar} \langle H \rangle - \frac{2}{\hbar} \langle H \rangle (\Psi,  \Psi) \\
& = & 0 \ .
\end{eqnarray*}	
\end{proof}

\section{Appendix C}
The reduced quantum optimization equation~(\ref{Eq2}) minimizes the system's total energy $\langle H \rangle$ (the expected value of $H$) and is guaranteed to converge.
\begin{proof}
\begin{eqnarray*}
\frac{d}{dt} \langle E(x) \rangle & = & \frac{d}{dt} (\Psi, E\Psi) \\
& = & \left(\frac{\partial \Psi}{\partial t}, E \Psi \right) + \left(\Psi, E \frac{\partial \Psi}{\partial t} \right)  \\
& = & \left(\frac{\partial}{\partial t} \left(\prod_i \Psi_i \right), E \Psi \right) + \left(\Psi, E \frac{\partial }{\partial t} \left(\prod_i \Psi_i \right)\right)  \\
& = & \sum_i \left(\frac{\partial \Psi_i}{\partial t}\Psi_{-i} , E \Psi_{-i} \Psi_i \right) + \left(\Psi_i \Psi_{-i}, E  \Psi_{-i}\frac{\partial \Psi_i}{\partial t} \right)   \\
& = & \sum_i \left(\frac{\partial \Psi_i}{\partial t}, E_i\Psi_i \right) + \left(\Psi_i, E_i\frac{\partial \Psi_i}{\partial t} \right)   \\
& = & \sum_i \left( -c (E_i -\langle E_i \rangle) \Psi_i, E_i\Psi_i \right) + \sum_i \left(\Psi_i, -c E_i(E_i -\langle E_i \rangle) \Psi_i \right) \\
& = &  -2c \sum_i \left( \langle E^2_i \rangle -\langle E_i \rangle^2 \right) \\
& \le & 0 \ .
\end{eqnarray*}	
The equality at the last step of the above proof holds true if and only if 
	the soft gradient descent reaches an equilibrium. 
That is, when $E_i$ satisfies 
\[ E_i \Psi_i = \langle E_i\rangle \Psi_i, \quad \mbox{for $i=1,2,\ldots, n$}. \]
This is exactly the time-independent Schr\"{o}dinger equation in quantum mechanics for the case of uncorrelated wavefunctions.

Therefore, The equation~(\ref{Eq2}) always converges to an equilibrium point defined by the time-independent Schr\"{o}dinger equation in quantum mechanics.

This completes the proof. $\S$
\end{proof}

\section{Appendix D}
The reduced quantum optimization equation~(\ref{Eq2}) defines a unitary transformation for $\Psi_i(x_i, t)$.
\begin{proof}
\begin{eqnarray*}
\frac{d}{dt} ||\Psi_i||^2 & = & \frac{d}{dt} ( \Psi_i, \Psi_i )\\
& = & \left(\frac{\partial \Psi_i}{\partial t}, \Psi_i \right) + \left(\Psi_i, \frac{\partial \Psi_i}{\partial t} \right)  \\
& = & \left( -c (E_i -\langle E_i \rangle) \Psi_i, \Psi_i \right) + \left(\Psi_i, -c(E_i -\langle E_i \rangle) \Psi_i \right) \\
& = &  -2c (\Psi_i, E_i \Psi) - 2c (\Psi_i, \langle E_i \rangle \Psi_i) \\
& = &  -2c \langle E_i \rangle - 2c \langle E_i \rangle (\Psi_i,  \Psi_i) \\
& = & 0 \ . \S
\end{eqnarray*}	
\end{proof}

\section{Appendix E}
The discrete time version for the reduced quantum optimization equation~(\ref{Eq3}) minimizes the system's total energy $\langle H \rangle$ (the expected value of $H$) and is guaranteed to converge.

\begin{proof}
Let $\Phi_1, \Phi_2, \ldots, \Phi_m$ are the eigen vectors of the local Hamiltonian $H_i$ with corresponding eigenvalues as 
$\lambda_1, \lambda_2, \ldots \lambda_m$ satisfying
\[ 0 \le \lambda_1 \le \lambda_2 \le \cdots \le \lambda_m. \]
Let the vector ${\bf \Psi}_i(t)$ be decomposed in terms of the eigen vectors as
\[ {\bf \Psi}_i(t) = \sum^{m}_{k=1} c_k \Phi_k \ . \]
Then we have
\begin{eqnarray*}
\langle H \rangle (t+1) & = & \left({\bf \Psi}(t+1), H{\bf \Psi} (t+1)\right) \\
& = & \left({\bf \Psi}_i(t+1) , H_i {\bf \Psi}_i(t+1) \right) \\
& = & \left(\frac{1}{Z_i} f(H_i)  {\bf \Psi}_i(t) , H_k \frac{1}{Z_i} f(H_i)  {\bf \Psi}_i(t) \right)  \\
& = & \frac{1}{\sum_k c^2_k f^2(\lambda_k)}\left(\sum_k c_k f(\lambda_k) \Phi_k, \sum_k c_k \lambda_k f(\lambda_k) \Phi_k \right) \\
& \le & \frac{1}{\sum_k c^2_k}\left(\sum_k c_k \Phi_k, \sum_k c_k \lambda_k \Phi_k  \right) \\
& = &\left({\bf \Psi}_i(t) , H_i {\bf \Psi}_i(t) \right)\\
& = & \langle H \rangle (t) 
\end{eqnarray*}	

The equality at the step 5 in the above proof holds true if and only if the iteration~(\ref{Eq3})
	reaches an equilibrium.
In this case, ${\bf \Psi}_i$ must be an eigenvector of the local Hamiltonian $H_i$ satisfying
\[ \lambda_i {\bf \Psi}_i = H_i {\bf \Psi}_i \ , \]
where $\lambda_i$ is the corresponding eigenvalue of the eigenvector.
It is exactly the time-independent Schr\"{o}dinger equation in quantum mechanics, just like the case of the soft gradient descent.

Therefore, the soft local search always converges to an equilibrium point defined by the time-independent Schr\"{o}dinger equation in quantum mechanics.
$\S$
\end{proof}

\end{document}